\documentclass[preprint,amsmath,amssymb,aps,showkeys,showpacs]{revtex4}

\usepackage[colorlinks=true, pdfstartview=FitV, linkcolor=blue, citecolor=red, urlcolor=magenta]{hyperref}
\usepackage{graphicx}
\usepackage{latexsym}
\usepackage{amsmath}
\usepackage{amsfonts}
\usepackage{amssymb}
\usepackage{verbatim}
\usepackage{dcolumn}
\usepackage{amssymb}
\usepackage{bm}
\usepackage{float}
\usepackage{subfigure}
\usepackage[english]{babel}
\graphicspath{{Figures/}}

\begin{document}
\newcommand{\be}{\begin{equation}}
\newcommand{\ee}{\end{equation}}
\newcommand{\bea}{\begin{eqnarray}}
\newcommand{\eea}{\end{eqnarray}}
\newcommand{\hf}{\frac{1}{2}}
\newcommand{\pa}{\partial}
\newcommand{\NITK}{
\affiliation{Department of Physics, National Institute of Technology Karnataka, Surathkal  575 025, India}
}

\newcommand{\IIT}{\affiliation{
Department of Physics, Indian Institute of Technology, Ropar, Rupnagar, Punjab 140 001, India
}}

\title{Dynamics and kinetics of phase transition for regular AdS black holes in general relativity coupled to non-linear electrodynamics }

\author{A. Naveena Kumara}
\email{naviphysics@gmail.com}
\NITK
\author{Shreyas Punacha}
\email{shreyasp444@gmail.com}
\NITK
\author{Kartheek Hegde}
\email{hegde.kartheek@gmail.com}
\NITK

\author{C. L. Ahmed Rizwan}
\email{ahmedrizwancl@gmail.com}
\NITK
\author{Md Sabir Ali}
\email{alimd.sabir3@gmail.com}
\IIT
\author{K. M. Ajith}
\email{ajith@nitk.ac.in}
\NITK

\begin{abstract}
Employing the free energy landscape, we study the phase transition and its dynamics for a class of regular black holes in Anti-de Sitter spacetime governed by the coupling of non-linear electrodynamics, which reduces to Hayward and Bardeen solutions for particular values of spacetime parameters. The Fokker-Planck equation is solved numerically by imposing the reflecting boundary condition and a suitable initial condition, using which, we investigate the probabilistic evolution of regular AdS black holes. In this approach, the on-shell Gibbs free energy is treated as a function of the radius of the event horizon, which happens to be the order parameter of the phase transition. The numerical solution is also obtained for the absorbing boundary condition. The dynamics of switching between the coexistence small black hole phase and large black hole phase due to the thermal fluctuation is probed by calculating the first passage time. The effect of temperature on the dynamical process is also investigated.

\end{abstract}


\maketitle


\section{Introduction}
One of the most intriguing aspects that ushered new research interests in recent years is the intrinsic connection between gravity and thermal systems. It is a well-established notion that black holes are not merely strong gravity objects but also possesses thermodynamic properties. In the seminal work of Hawking and Bekenstein \citep{Hawking:1974sw, Bekenstein1973}, the temperature and entropy were defined for a black hole system, which on the gravity side correspond to the area of the event horizon and surface gravity, respectively. Moreover, the later developments show that the black holes exhibit a rich class of phase transitions analogous to real-life thermodynamic phenomena. In a specific scenario, the black hole thermodynamics gained more attention, where the longstanding question about pressure and volume of the black hole was discussed in the context of black hole thermodynamics \citep{Kastor:2009wy, Dolan:2011xt}. The thermodynamic variable pressure was introduced via dynamical cosmological constant and volume from its conjugate quantity. With this extended phase space, charged AdS black holes exhibit phase structure similar to that of van der Waals fluid \citep{Kubiznak2012, Gunasekaran2012, Kubiznak:2016qmn}. A first-order transition between a small black hole (SBH) phase and a large black hole phase (LBH) is observed analogous to a liquid-gas transition in conventional thermodynamics. 

A new step forward in understanding the black hole phase transition is the introduction of a free energy landscape aided with stochastic Fokker-Planck equation to analyse the kinetics and dynamics of the transition process \citep{PhysRevD.102.024085}. In the context of black hole physics, the idea was originally developed to uncover the kinetics of the Hawking-Page transition. Later it was generalised to charged AdS black hole, where the transition between a metastable state and a stable state was analysed \citep{Li:2020nsy}. The dynamics of the specific case of transition along the coexistence line is studied for the five-dimensional Gauss-Bonnet black hole in Ref. \citep{Wei:2020rcd}. Here, the transition is between the stable SBH and LBH states. The results show that the system can make a transition from the initial state to other stable state and switch back. Soon, the formulation has been discussed in different contexts \citep{Li:2021zep, Li:2021vdp, Xu:2021qyw}. The dynamic property of triple point \citep{Wei:2021bwy}, effect of dark energy \citep{Lan:2021crt}, in Kerr-AdS spacetime \citep{Yang:2021nwd}, in charged dilaton black holes \citep{Mo:2021jff}, four-dimensional Gauss-Bonnet black hole \citep{Li:2020spm} are the further applications of the free energy landscape in black hole physics. It is important to stress that the off-shell Gibbs free energy plays a vital role in these investigations, which is the driving force for the black hole phase transition and it is treated as a function of the horizon radius, which is the order parameter of the transition. 

In general relativity, the physical singularity at the centre of the black hole exists due to the Penrose and Hawking theorems \citep{Hawking:1969sw, Hawking:1973uf}. However, there are several methods to bypass this singularity and to obtain singularity free black hole solutions. One of such interesting class of such solutions is governed by the coupling of non-linear electrodynamics \citep{Fan:2016hvf, Toshmatov:2018cks}. In these solutions, the spacetime is free from a singularity when the mass parameter is solely obtained from a non-linear electrodynamics source. In fact, these regular solutions are often interpreted as the gravitational field of a non-linear electric or magnetic monopole. This formulation has special cases which are important in the literature, namely the Bardeen and Hayward solutions \citep{bardeen1968non, Hayward:2005gi}. These black holes in AdS spacetime exhibit phase structure similar to van der Waals fluid \citep{Fan:2016rih, Tzikas:2018cvs}. In the present work, we investigate the dynamics of these phase transitions using the Gibbs free energy landscape.

The organisation of the paper is as follows. In the next section (\ref{haywardsection}), we present the thermodynamics and phase transition of the Hayward solution. In the same section, we probe the dynamics of that transition using Gibbs free energy landscape. In section  \ref{bardeensecion} we carry out a similar investigation for the Bardeen solution. The results are summarised in section  \ref{discussions} with discussions.

\section{Hayward AdS Black Hole}
\label{haywardsection}
\subsection{Thermodynamics and phase transition}

We consider the black hole solutions derived from Einstein gravity minimally coupled to nonlinear electrodynamics with negative cosmological constant $\Lambda$, which are governed by the action \cite{Fan:2016hvf},
\begin{equation}
\label{action}
    \mathcal{I}=\frac{1}{16\pi G} \int{d^4x}\sqrt{-\hat{g}}[R-\mathcal{L\left(F\right)}+2\Lambda].
\end{equation}
In the above equation, $R$ and  $\hat{g}$ are the Ricci scalar and the determinant of the metric tensor, respectively. The Lagrangian density $\mathcal{L(F)}$ is a function of $\mathcal{F} = F_{\mu \nu }F^{\mu \nu }$ with $F_{\mu\nu}=2\nabla_{[\mu}A_{\nu]}$, the field tensor of nonlinear electrodynamics. The Hayward-AdS black hole spacetime is due to the Lagrangian density,
\begin{eqnarray}
\label{lagran}
\mathcal{L\left(F\right)}=\frac{12}{\alpha}\frac{\left(\alpha \mathcal{F}\right)^{3/2}}{\left(1+\left(\alpha \mathcal{F}\right)^{3/4}\right)^2},
\end{eqnarray}
with $\alpha >0$ which has the dimension of length squared. The non-vanishing components of $F_{\mu\nu}$ in a spherically symmetric spacetime are $F_{tr}$ and $F_{\theta\phi}$. For the case of pure magnetic charge, only $F_{\theta\phi}$ survives. Then the corresponding Maxwell tensor has only one non-zero component,
\begin{eqnarray}\label{maxwell}
F_{\theta\phi}=-F_{\phi\theta}=-Q_m\sin\theta.
\end{eqnarray}
It is easy to read the gauge potential and the Maxwell invariant for this field as,
\begin{eqnarray}
\label{gauge}
A_\mu=Q_m\cos\theta\delta_\mu^\phi, \qquad \mathcal{F}=\frac{2 Q_m^2}{r^4}.
\end{eqnarray}
The constant $Q_m$ is the magnetic monopole charge of the nonlinear electrodynamics. The metric for the Hayward AdS spacetime in four-dimension reads \cite{Fan:2016hvf},
\begin{eqnarray}\label{metric2}
\mathrm{ds^2}=-f(r)dt^2+\frac{1}{f(r)}dr^2+r^2 d\Omega_{2}^2,
\end{eqnarray}
where $d\Omega_{2}^2=d\theta^2+\sin^2\theta{d\phi^2}$, is a $2$-dimensional unit sphere, and the corresponding  metric function is given by,
\begin{equation}
    f(r)=\left(1-\frac{2 M r^2}{r^3+g^3}-\frac{\Lambda r^2}{3}\right),
\end{equation}
in which $M$ is the mass of the black hole and $g$ the parameter related to the magnetic charge $Q_m$, as $Q_m=g^2/\sqrt{2\alpha}$. 

Now, we proceed to discuss the extended thermodynamics of the black hole, where the pressure $P$ is related to the dynamic cosmological constant $\Lambda$ as \citep{Kastor:2009wy, Dolan:2011xt},
\begin{equation}
   P= -\frac{\Lambda}{8\pi}.  
\end{equation}
In this description, the mass $M$ of the black hole is treated as enthalpy rather than energy. It can be expressed in terms of horizon radius $r_+$, by using the condition, $f(r_+)=0$,
\begin{equation}
    M=\frac{r_+}{2}+\frac{4}{3} \pi   P (g^3+r_+^3)+\frac{g^3}{2 r_+^2}.
\end{equation}
The Hawking temperature $T$ is associated with the surface gravity $\kappa$, which is obtained as,
\begin{align}
T&=\frac{\kappa}{2\pi}=\left. \frac{f'(r)}{4\pi} \right|_{r=r_+}\nonumber \\
&=\frac{2 P r_+^4}{g^3+r_+^3}-\frac{g^3}{2 \pi  r_+ \left(g^3+r_+^3\right)}+\frac{r_+^2}{4 \pi  \left(g^3+r_+^3\right)}.
\label{teqn}
\end{align}
The first law of thermodynamics can be written as,
\begin{equation}
dM=TdS+\Psi dQ_m+VdP+\Pi d \alpha,
\end{equation}
where $\Psi$ and $\Pi$ are the conjugate variables corresponding to the magnetic charge $Q_m$ and parameter $\alpha$, respectively. Unlike the charged AdS black holes, the entropy  and volume of the Hayward black hole have the following modified form,
\begin{align}
S&=\int \frac{dM}{T}=2 \pi  \left(\frac{r_+^2}{2}-\frac{g^3}{r_+}\right),\\
V&=\left( \frac{\partial M}{\partial P}\right)_{S,Q_m,\alpha}=\frac{4}{3} \pi  \left(g^3+r_+^3\right).
\end{align}
The equation of state of the system is,
\begin{equation}
P=\frac{g^3}{4 \pi  r_+^5}+\frac{g^3 T}{2 r_+^4}-\frac{1}{8 \pi  r_+^2}+\frac{T}{2 r_+}.
\end{equation}
The Hayward black hole exhibits a vdW like phase transition between a large black hole phase (LBH) and small black hole phase (SBH) \citep{Fan:2016rih}. The critical point of this first order phase transition is governed by the monopole charge parameter $g$, which is given by,
\begin{equation}
T_{cH}=\frac{\left(5 \sqrt{2}-4 \sqrt{3}\right) \left(3 \sqrt{6}+7\right)^{2/3}}{4\times 2^{5/6} \pi  g},
\end{equation}
\begin{equation}
P_{cH}=\frac{3 \left(\sqrt{6}+3\right)}{16\times 2^{2/3} \left(3 \sqrt{6}+7\right)^{5/3} \pi  g^2}.
\end{equation}


\subsection{Gibbs free energy landscape}
\label{sec3}
The signature of the first-order phase transition of the black hole can be clearly seen in the swallowtail behaviour of the Gibbs free energy. Therefore, free energy is a powerful tool to investigate the thermal dynamic phase transition of the black hole. For Hayward AdS black hole, the Gibbs free energy is,
\begin{equation}
    G=M-T S=\frac{8 g^3 r_+^3 \left(10 \pi  P r_+^2+3\right)+2 g^6 \left(8 \pi  P r_+^2-3\right)-8 \pi  P r_+^8+3 r_+^6}{12 r_+^2 \left(g^3+r_+^3\right)}.
    \label{gibbs1}
\end{equation}
The swallowtail behaviour appears for the pressure below the critical value $P_{cH}$ and the first derivative of $G$ is discontinuous in such conditions as it corresponds to a first order phase transition. The behaviour disappears at the critical pressure  where the first derivative becomes continuous and the corresponding transition is second order. 

\begin{figure*}[!ht]
\centering
\subfigure[ref2][]{\includegraphics[scale=0.8]{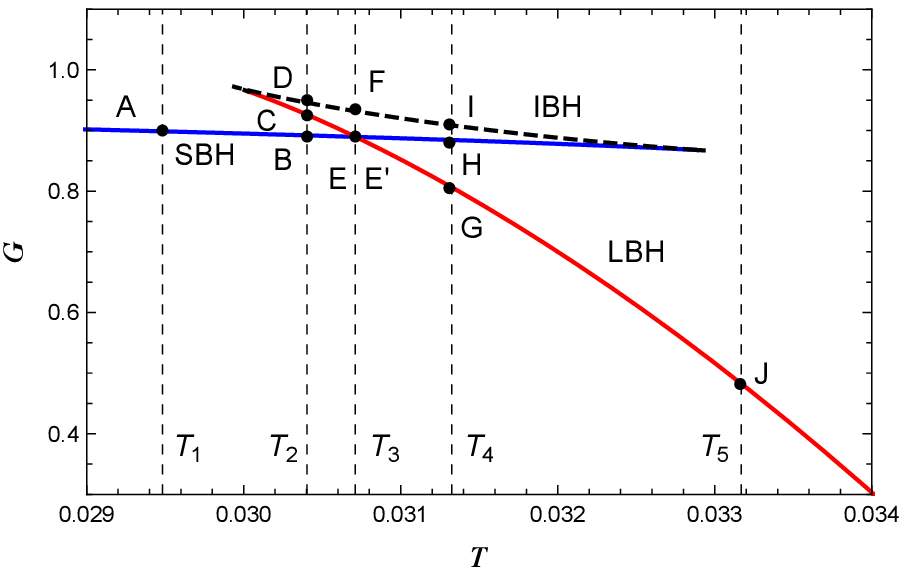}
\label{gt}}
\qquad
\subfigure[ref3][]{\includegraphics[scale=0.8]{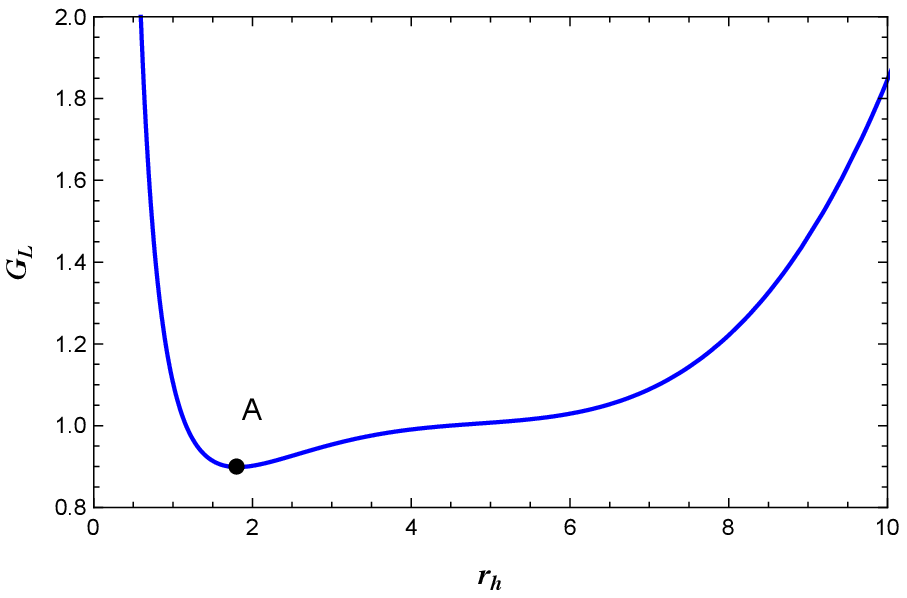}
\label{gr1}}
\subfigure[ref2][]{\includegraphics[scale=0.8]{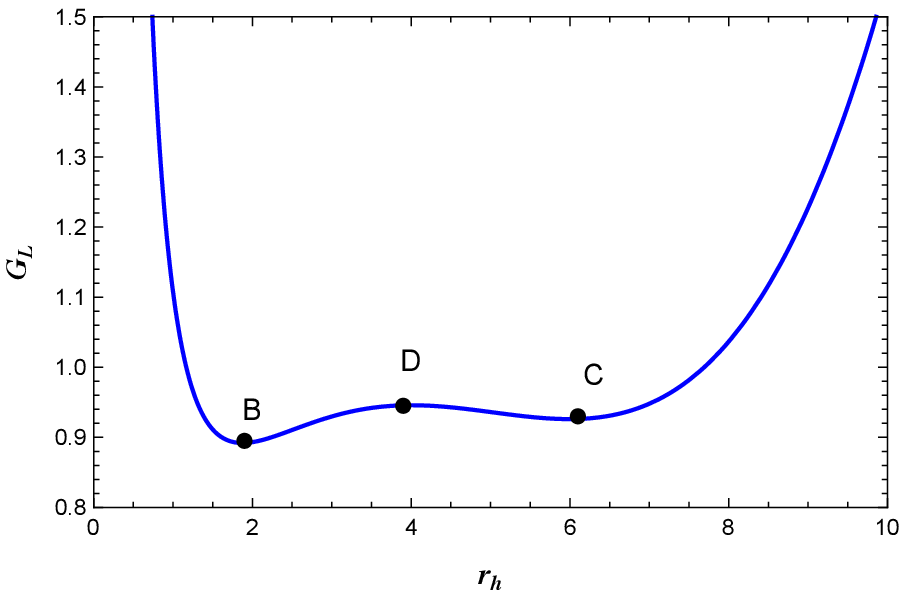}
\label{gr2}}
\qquad
\subfigure[ref3][]{\includegraphics[scale=0.8]{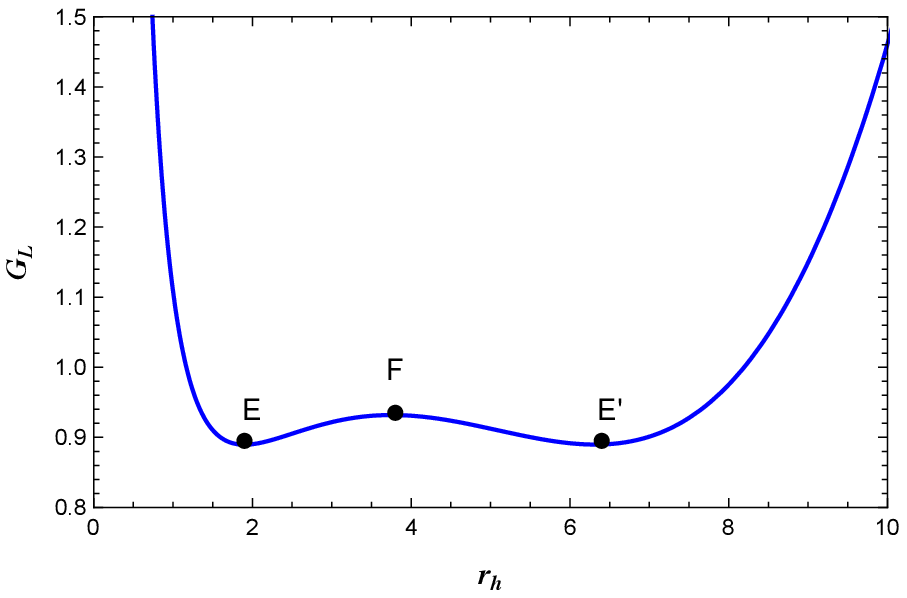}
\label{gr3}}
\subfigure[ref2][]{\includegraphics[scale=0.8]{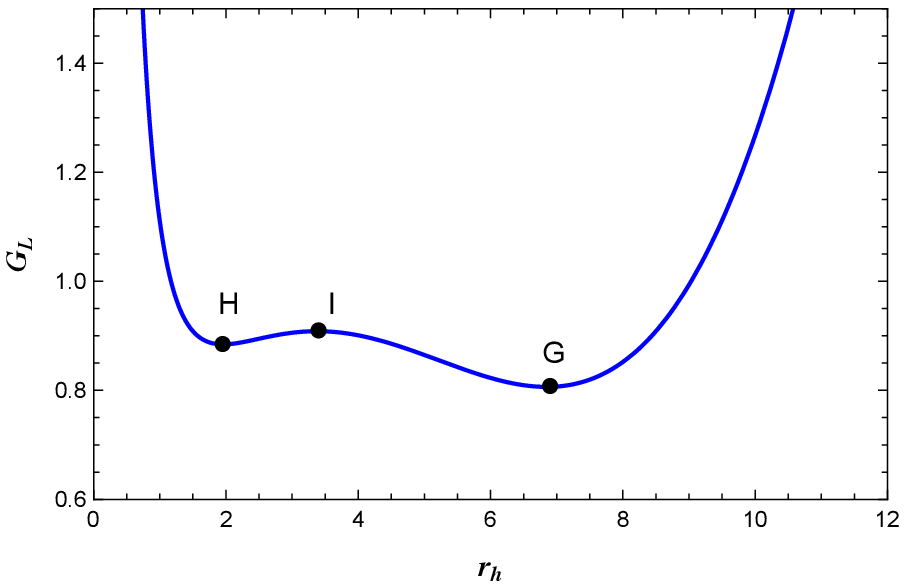}
\label{gr4}}
\qquad
\subfigure[ref3][]{\includegraphics[scale=0.8]{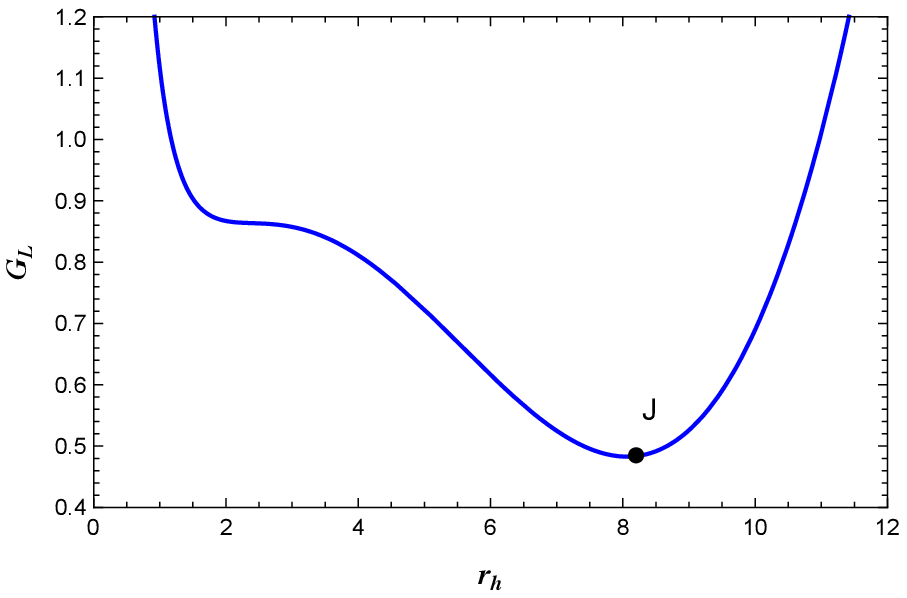}
\label{gr5}}
\caption{The Gibbs free energy plots with $g=1$ and $P=0.06P_c$. (a) The swallowtail behaviour of Gibbs free energy, $G$ vs $T$ plot. The solid red, solid blue and dashed black lines correspond to SBH, LBH and IBH phases, respectively. (b)-(f) are the off-shell Gibbs free energy $G_L$ vs $r_h$ plots. The ensemble temperatures are taken as (b) $T=0.96T_{co}$, (c) $T=0.99T_{co}$, (d) $T=T_{co}$, (e) $T=1.02T_{co}$ and (f) $T=1.08T_{co}$, where $T_{co}=0.03071$ is the coexistence temperature corresponding to the pressure  $P=0.06P_cH$.}
\label{gibbsplot}
\end{figure*}

In the free energy landscape, we consider a canonical ensemble with a fixed temperature $T_E$. The black hole states in this ensemble have a different radius at the specified temperature. In the equilibrium state, the black hole temperature is identical to the ensemble temperature. In the free energy landscape, the Gibbs free energy is expressed as,
\begin{equation}
    G_L=M-T_E S=\frac{g^3 \left(8 \pi  P r_+^2+12 \pi  r_+ T+3\right)+r_+^3 \left(8 \pi  P r_+^2-6 \pi  r_+ T+3\right)}{6 r_+^2}.
    \label{gibbs2}
\end{equation}
The behaviour of Gibbs free energy with the horizon radius for a fixed value of pressure and temperature gives information about the SBH-LBH first-order transition. We study this behaviour in Fig \ref{gibbsplot} using the Eqns. \ref{gibbs1} and \ref{gibbs2}. The swallowtail behaviour is displayed in Fig. \ref{gt}, which gives the coexistence temperature $T_0$ for a given coexistence pressure $P_0$. We seek this method to obtain the coexistence temperatures for different coexistence pressures, as the analytical expression for the coexistence curve is not always feasible for complicated black hole spacetimes. In fact, one can obtain a numerical fit for the coexistence curve using this method. Here we have taken $P=0.6 P_{cH}$. As the heat capacity governs the stability of the black hole phase, the solid blue and solid red lines with positive specific heat represents stable SBH and LBH phases, respectively, whereas, the dashed black line with negative specific heat corresponds to the unstable intermediate black hole phase. For a fixed pressure $P=0.6 P_{cH}$, the system evolves with an increase in the temperature, through the path $A-B-E-G-J$, by choosing the least values of Gibbs free energy. The stable LBH below $T_3$ and stable $SBH$ above $T_3$ are metastable states which are globally unstable.  

In the next series of plots, the Gibbs free energy is plotted as a function of horizon radius, which is the order parameter for the transition. Not all points in the $G_L-r_+$ curve correspond to the physical states of black holes; rather, only the extremal points do. In this meaning, $G_L$ is termed as off-shell Gibbs free energy or generalised Gibbs free energy. For temperature $T_1$, there is only one physical black hole state ($A$ in Fig. \ref{gr1}), which is an SBH state. At temperature $T_2$ there exist three extremum, $B$ , $C$, and $D$ (Fig. \ref{gr2}). The local maximum $D$ is an unstable intermediate black hole state. The local minimum $B$ and $C$ correspond to the stable SBH  and a metastable SBH state, respectively. As the system prefers a lowest Gibbs free energy state, the black hole will choose state $B$ at this temperature. As the temperature is increased to a value $T_3$, the depth of the basins become equal, and both ($E$ and $E'$) correspond to stable states with equal priority. This is the coexistence temperature, and the scenario corresponds to the first order SBH-LBH phase transition. Further, an increase in temperature to $T_4$ results in a stable LBH state at $G$ and a metastable SBH state at $H$. At temperature $T_5$, the system has only one physical state, which is the stable LBH at $J$. The value of $r_+$ at these basins correspond to the respective horizon radius of the SBH or LBH phases at every temperature. As the physical states of the black hole correspond to the vanishing first derivative of $G_L$, from Eq. \ref{gibbs2} we have,
\begin{equation}
   \frac{1}{2} -2 \pi  T \left(\frac{g^3}{r_{hs}^2}+r_{hs}\right)-\frac{g^3}{r_{hs}^3}+4 \pi  P r_{hs}^2=0,
\end{equation}

\begin{equation}
   \frac{1}{2} -2 \pi  T \left(\frac{g^3}{r_{m}^2}+r_{m}\right)-\frac{g^3}{r_{m}^3}+4 \pi  P r_{m}^2=0.
   \label{rmeq}
\end{equation}

\begin{equation}
   \frac{1}{2} -2 \pi  T \left(\frac{g^3}{r_{hl}^2}+r_{hl}\right)-\frac{g^3}{r_{hl}^3}+4 \pi  P r_{hl}^2=0.
\end{equation}
The value of horizon radii  $r_{hs}$ of SBH phase, $r_{hl}$ of LBH phase and $r_m$ of the intermediate phase can be obtained by using these conditions. For a given pressure, the double basin behaviour of $G_L$ can be seen in a range of temperature $T_{min}<T<T_{max}$ which can be determined by using the condition $\partial  T/\partial r_+=0$ in Eq. \ref{teqn}. The basins correspond to stable states, which are separated by a finite-height barrier of the intermediate phase. However, in the ensemble perspective, one stable state can transit into another stable state due to thermal fluctuations. The dynamics of this probabilistic evolution of the system between a SBH and LBH phase is characterised by the Fokker-Planck equation.


\subsection{Dynamic properties of phase transition}
In this section we will study the dynamics of the thermodynamic phase transition of the black hole for the case of SBH-LBH transition, which corresponds to double wells with the same depth in free energy landscape. In the following section we use $r$ for $r_+$ for simplicity.
\subsubsection{Fokker-Planck equation and probabilistic evolution}
A systematic analysis of thermodynamics and kinetics of black holes from Gibbs free energy landscape can be done using Fokker Planck equation \citep{Li:2020nsy},
\begin{equation}
    \frac{\partial \rho (r,t)}{\partial t}=D\frac{\partial }{\partial r} \left( e^{-\beta G_L(T,P,r)} \frac{\partial }{\partial r} \left( e^{\beta G_L(T,P,r)} \rho (r,t)\right) \right),
\end{equation}
where $\beta =1/k_BT$ with $k_B$ the Boltzmann constant, $D$ is diffusion constant which is given by $D=k_BT/\xi$ with $\xi$ as dissipation coefficient and $\rho (r,t)$ the probability distribution of black hole states. We impose the following reflection and absorbing boundary conditions (at $r=r_0$) to solve the Fokker-Planck equation, 
\begin{equation}
    \beta \partial _r G_L(T,P,r_0)\rho (r_0,t)+\partial _r \rho(r_0,t)=0 \qquad \text{and} \qquad  \rho(r_0,t)=0,
\end{equation}
and the initial condition,
\begin{equation}
    \rho (r,0)=1/(\sqrt{\pi} a) e^{-(r-r_i)^2/a^2},
\end{equation}
where $r_i$ are the radii of small and large black hole phases. We take $r_i$ as $r_{hs}$ and $r_{hl}$ as we are focused on the transition between SBH and LBH at the coexistent temperature. Here, $a$ is a constant which determines the initial width of the probability distribution wave packets, which does not influence the final results. Initially the black hole is chosen to be in any one state, either SBH or LBH state. As the phase transition is possible at coexistent temperature, it is expected that, with time evolution, there will be nonzero probability for both states. In fact, this probability evolution due to fluctuating macroscopic variables is governed by the Fokker-Planck equation. The stochastic fluctuating variable for the black hole phase transition is the order parameter horizon radius. To be precise, the above equation is a special case of Fokker-Planck equation known as the Smoluchowski equation.

\begin{figure*}[tbh]
\centering
\subfigure[ref2][]{\includegraphics[scale=0.8]{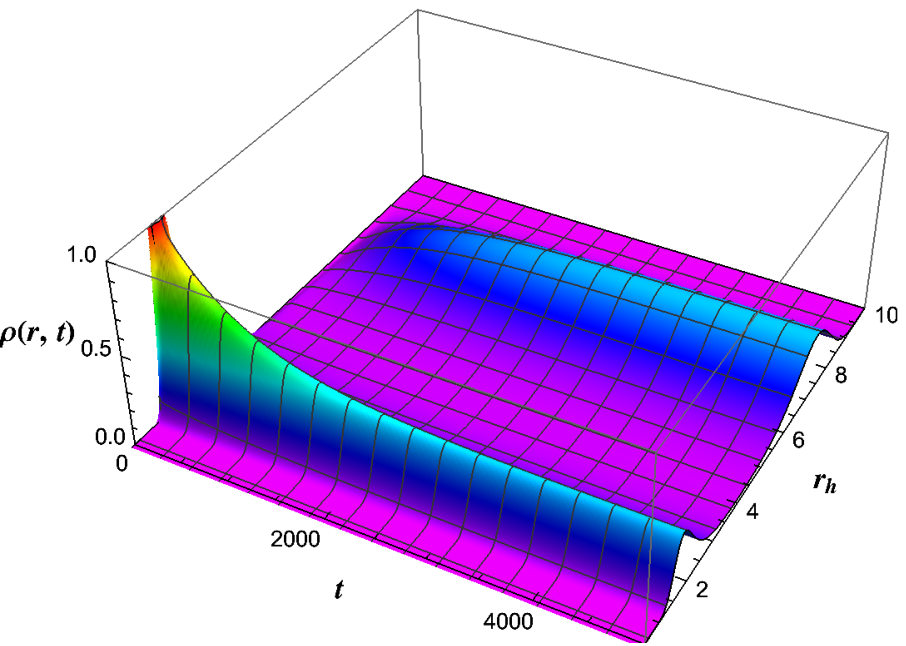}
\label{Hrbcsbh3dt1}}
\qquad
\subfigure[ref3][]{\includegraphics[scale=0.8]{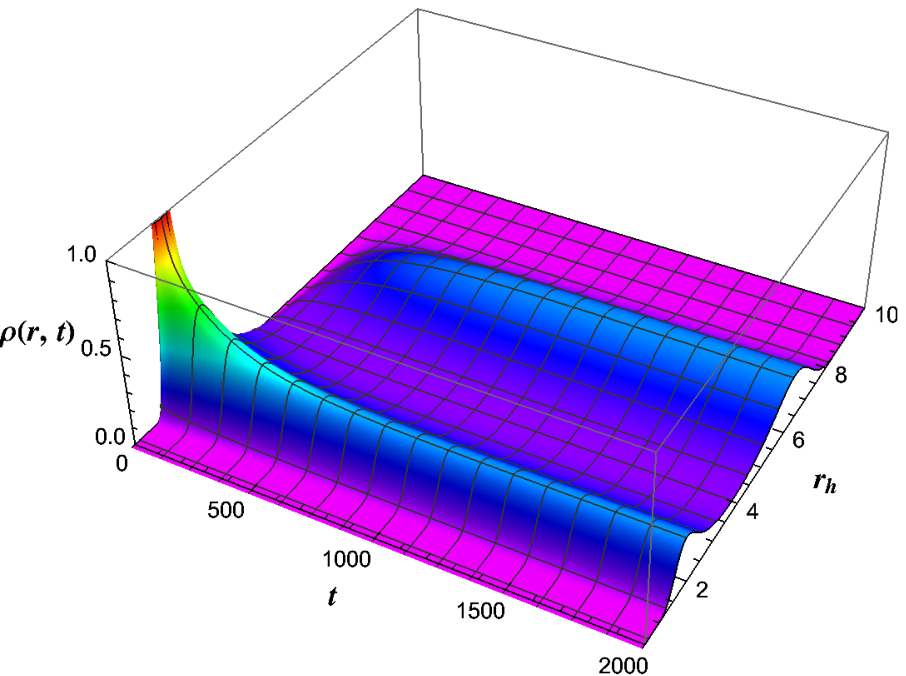}
\label{Hrbcsbh3dt2}}
\subfigure[ref2][]{\includegraphics[scale=0.8]{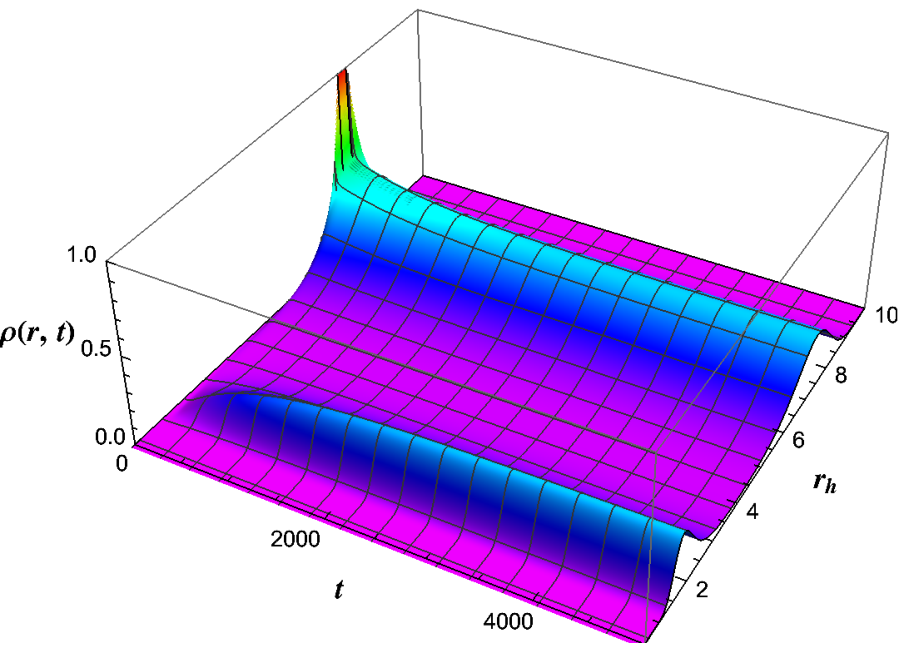}
\label{Hrbclbh3dt1}}
\qquad
\subfigure[ref3][]{\includegraphics[scale=0.8]{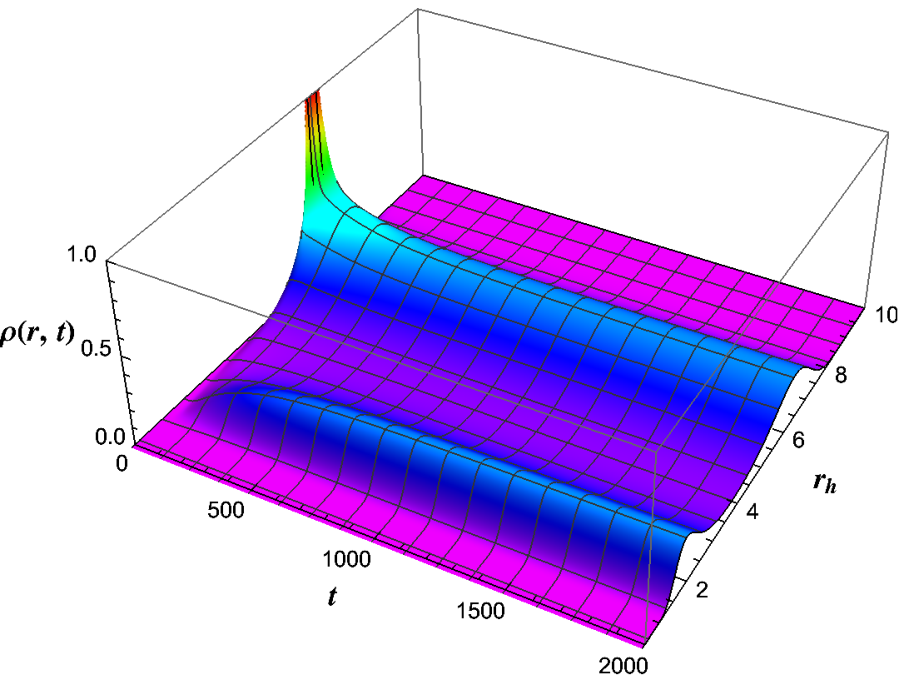}
\label{Hrbclbh3dt2}}
\caption{The time evolution of the probability distribution $\rho (r, t)$.
In (a) and (b) the initial Gaussian wave packet is placed at SBH state and in (c) and (d) they are located at LBH state. The reflection boundary conditions are imposed at $r=0$ and $r=\infty$. The coexistent temperatures correspond to $P=0.5P_{cH}$ ($T_E=0.0284$ in left panel) and  $P=0.6P_{cH}$ ($T_E=0.0307$ in right panel) with $g=1$.}
\label{H_time_evolution}
\end{figure*}

\begin{figure*}[tbh]
\centering
\subfigure[ref2][]{\includegraphics[scale=0.8]{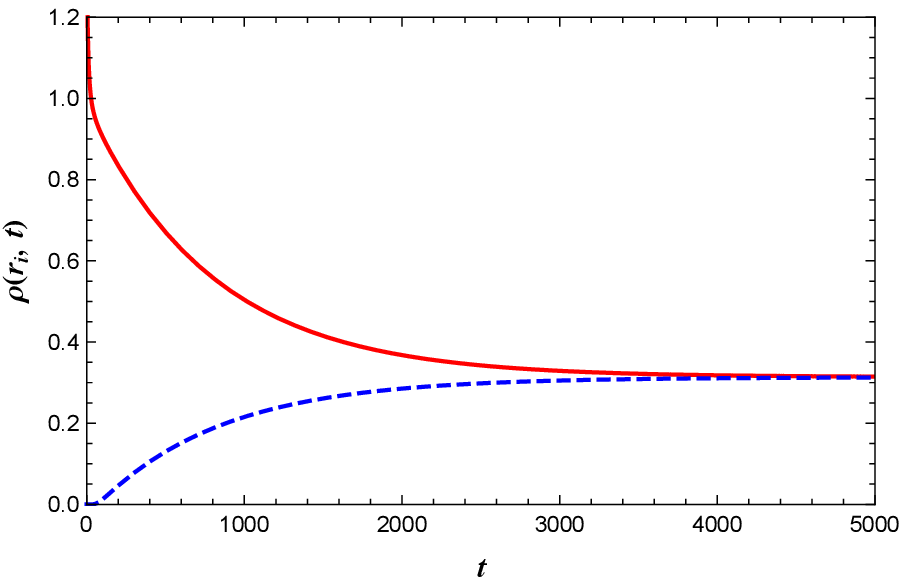}
\label{Hrbcsbh2dt1}}
\qquad
\subfigure[ref3][]{\includegraphics[scale=0.8]{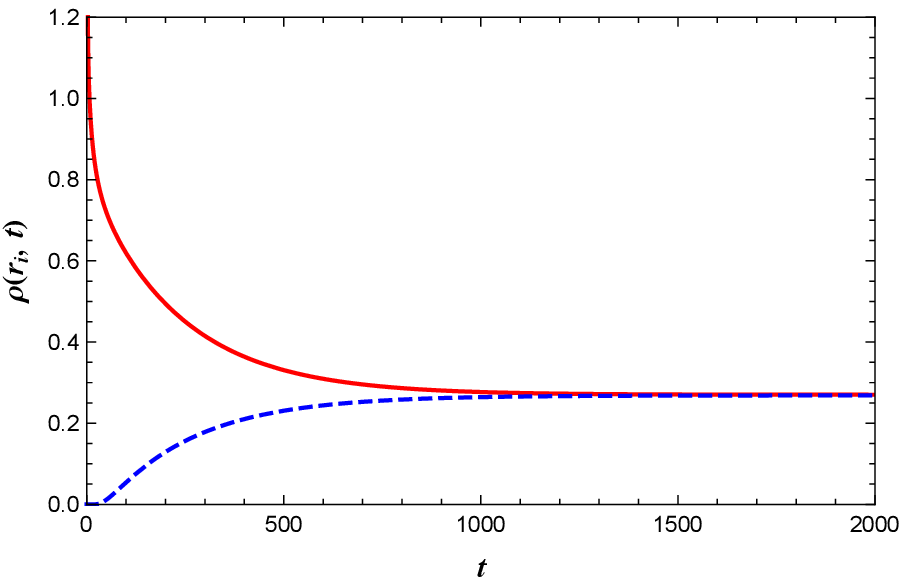}
\label{Hrbcsbh2dt2}}
\caption{The time evolution of the probability distribution function $\rho (r,t)$. The solid red and dashed blue curves correspond to $\rho (r_{hs},t)$ and $\rho (r_{hl},t)$, respectively. The initial wave packet is located at SBH state. The coexistent temperatures are (a) $T_E=0.0284$. (b) $T_E=0.0307$, with $g=1$.}
\label{H_time_evolution_2d}
\end{figure*}

First, we consider the reflection boundary condition, which preserves the normalisation of probability distribution $\rho (r,t)$, with the boundaries located at $r=0$ and $r=\infty$. The time evolution of the probability distribution for different initial conditions and temperatures are depicted in Fig. \ref{H_time_evolution}. We have considered initial wave packet at SBH state by choosing $r=r_{hs}$ and at LBH by letting $r=r_{hl}$ for two different coexistent temperatures corresponding to $P=0.4 P_{cH}$ and $P=0.6P_{cH}$. We set the initial width of the Gaussian wave packet for $a=0.1$ for the convenience of numerical calculation. In Figs. \ref{Hrbcsbh3dt1} and \ref{Hrbcsbh3dt2} initially the black hole is kept in SBH phase $(t=0)$. As the temperature increases, the wave packet spreads to the LBH state, correspondingly reducing the probability distribution in the SBH state. The distribution quickly becomes quasi-stationary with two peaks representing two wells in $G_L-r_+$ plots. As $t\rightarrow \infty$, the distribution becomes stationary with both peaks having equal weight. Similar phenomena with leakage from LBH to SBH is observed in Figs. \ref{Hrbclbh3dt1} and \ref{Hrbclbh3dt2} where the initial distribution was localised to $r=r_{hl}$. 

This evolution of probability density of leaking from the initial state to other can be made more apparent by looking at $\rho (r_{hs},t)$ and $\rho (r_{hs},t)$, where we examine the height of the wave packets the global minima of the Gibbs free energy. For the initial state at SBH, the behaviour is shown in Fig. \ref{H_time_evolution_2d}. For $t=0$, the probability density $\rho$ is maximum for the SBH branch, whereas, it is zero for the LBH branch. Later in the limit $t\rightarrow \infty$, both approaches the same value. For higher ensemble temperature, this saturation of distribution is attained quickly (relaxation time is small). This is due to an increase in thermal fluctuation with the increase in temperature, which prompts the leakage of states in the ensemble to reach equilibrium distribution. This can also be seen in the decreasing height of the potential barrier between SBH and LBH in the Gibbs free energy plots.

\subsubsection{The first passage time}
An important quantity in LBH-SBH phase transition, which is a stochastic dynamic process, is the first passage time. As the process is due to the thermal fluctuation in the ensemble, the first passage time is a random variable. It is defined as the time taken by the initial black hole state to reach the unstable intermediate state for the first time. In other words, it is the time required to climb the potential barrier of free energy from the stable LBH or SBH state for the first time. The mean of this quantity is the measure of the timescale of the phase transition.

\begin{figure*}[tbh]
\centering
\subfigure[ref2][]{\includegraphics[scale=0.8]{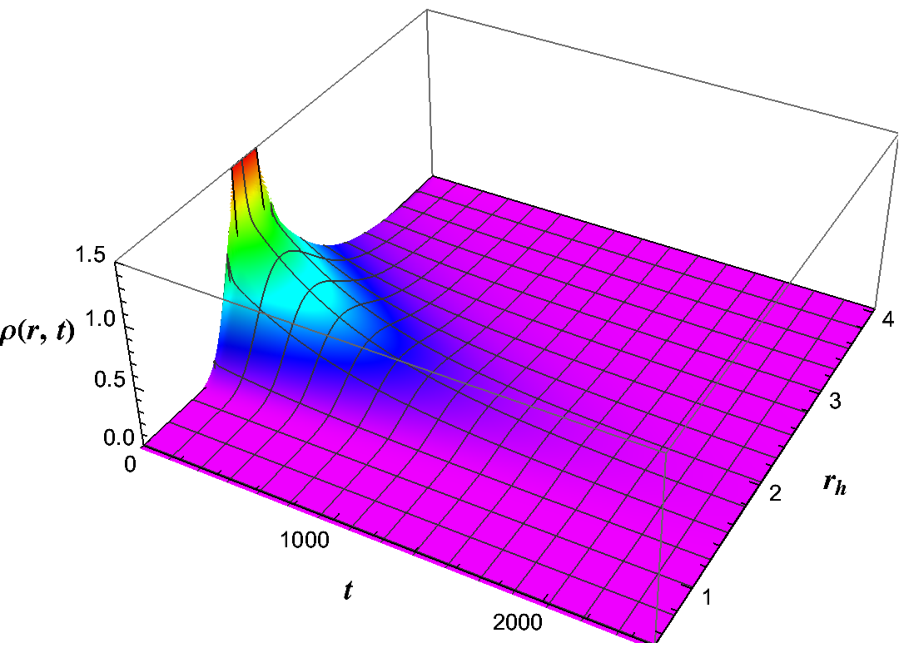}
\label{Habcsbh3dt1}}
\qquad
\subfigure[ref3][]{\includegraphics[scale=0.8]{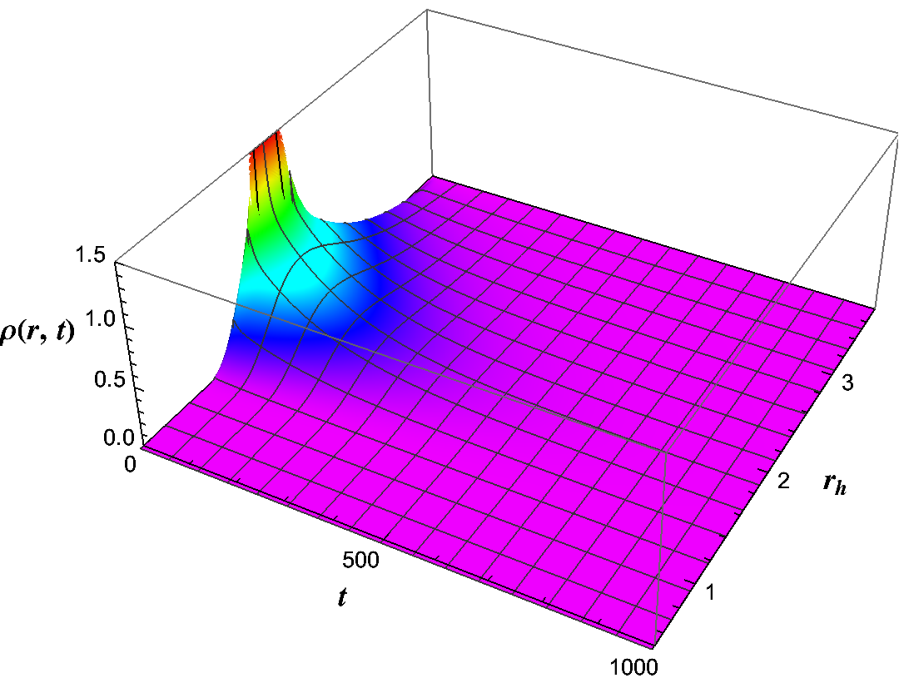}
\label{Habcsbh3dt2}}
\subfigure[ref2][]{\includegraphics[scale=0.8]{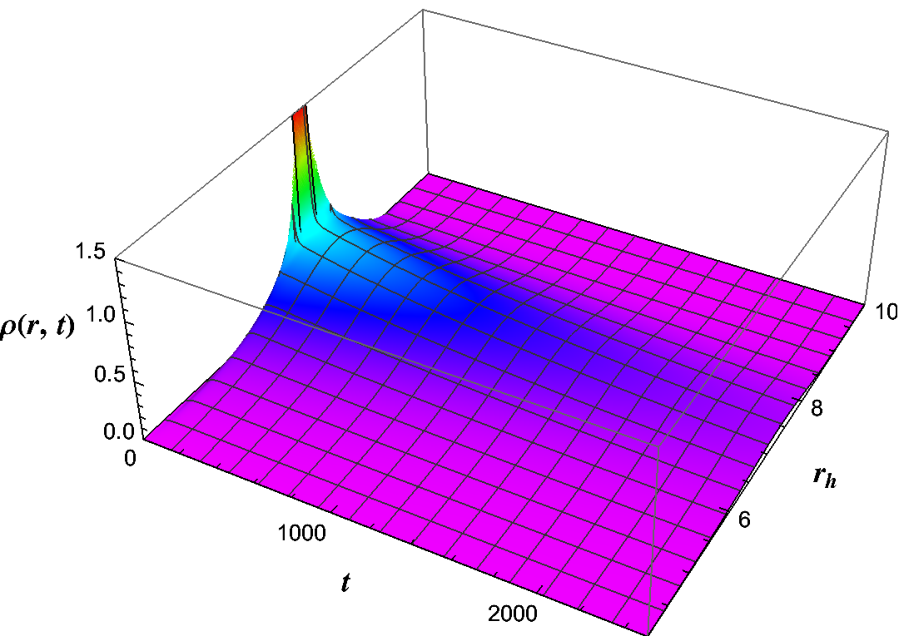}
\label{Habclbh3dt1}}
\qquad
\subfigure[ref3][]{\includegraphics[scale=0.8]{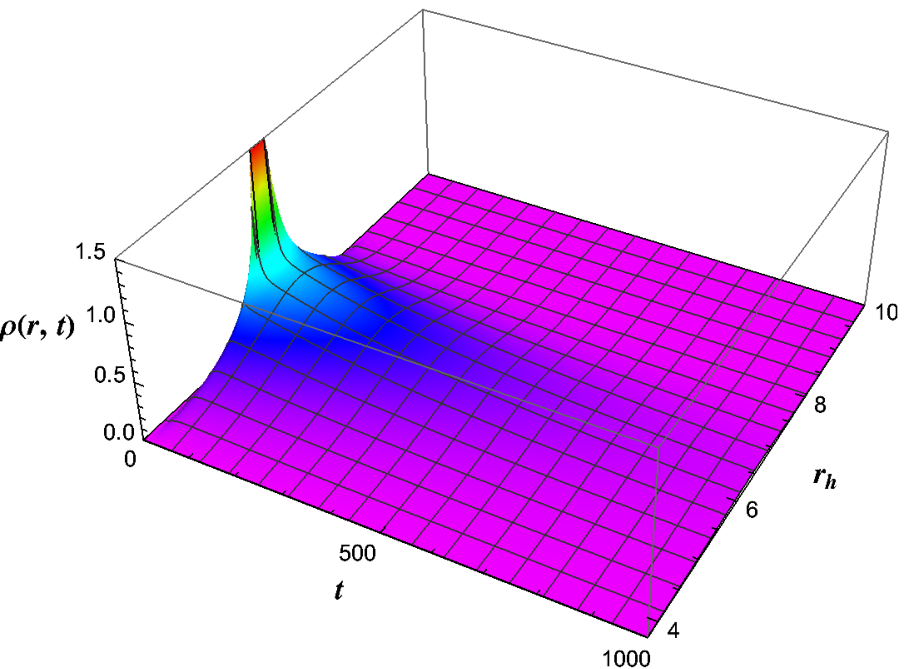}
\label{Habclbh3dt2}}
\caption{Plots depicting the time evolution of the probability distribution $\rho (r,t)$ with $g=1$. In (a) and (b) initial Gaussian wave packet is at SBH and in (c) and (d) it is at LBH states. The absorbing boundary condition is imposed at $r=r_m$ and reflecting boundary condition at other boundaries. The coexistent temperatures are (a) $T_E=0.0284$. (b) $T_E=0.0307$. (c) $T_E=0.0284$. (d) $T_E=0.0307$, with $g=1$.}
\label{H_time_evolution_absorbing}
\end{figure*}

\begin{figure*}[tbh]
\centering
\subfigure[ref2][]{\includegraphics[scale=0.8]{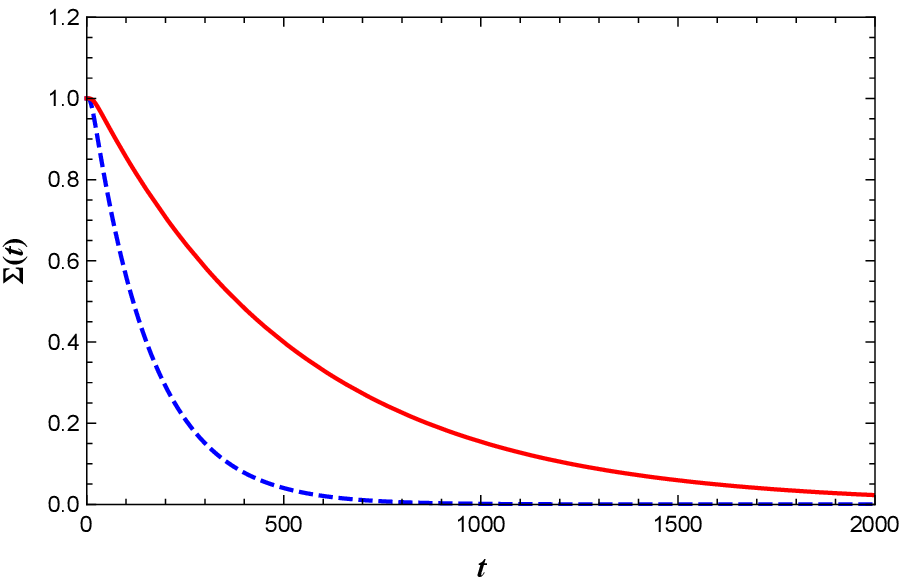}
\label{probsbh}}
\qquad
\subfigure[ref3][]{\includegraphics[scale=0.8]{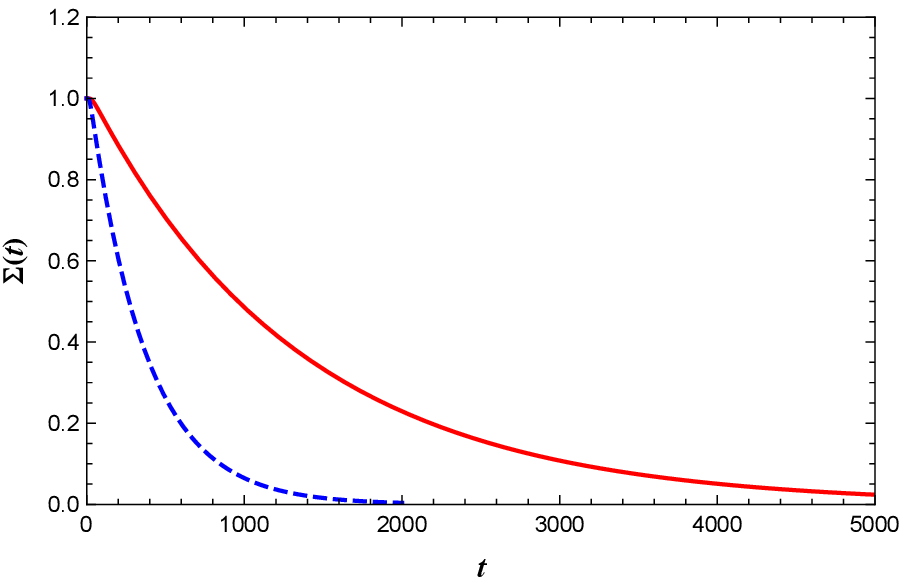}
\label{problbh}}
\caption{The time evolution of the probability distribution $\Sigma (t)$ that the system stays at the initial state. (a) Initial SBH and (b) initial LBH state. Red solid and blue dashed curves are for the coexistent temperatures $T_E=0.0284$ and  $T_E=0.0307$, respectively, with $g=1$.}
\label{prob}
\end{figure*}

\begin{figure*}[tbh]
\centering
\subfigure[ref2][]{\includegraphics[scale=0.8]{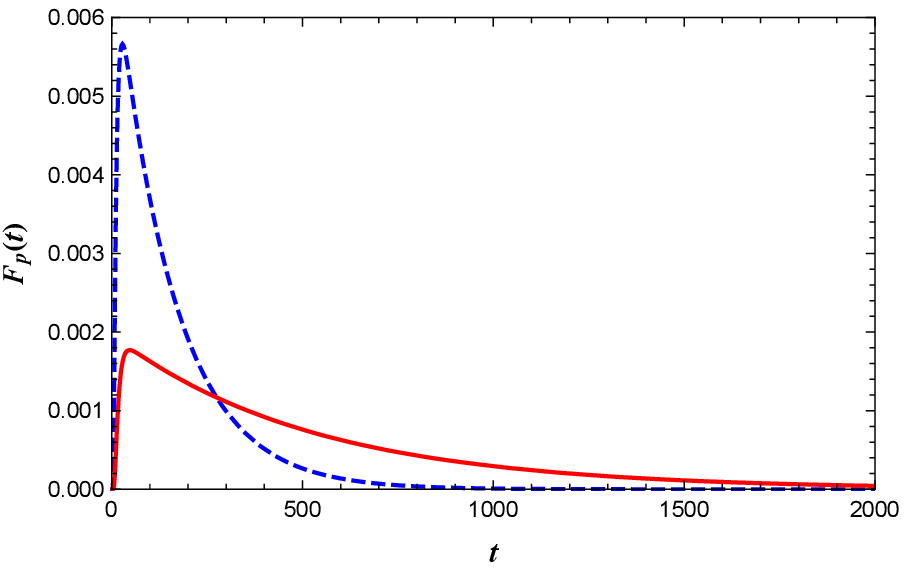}
\label{firstpasssbh}}
\qquad
\subfigure[ref3][]{\includegraphics[scale=0.8]{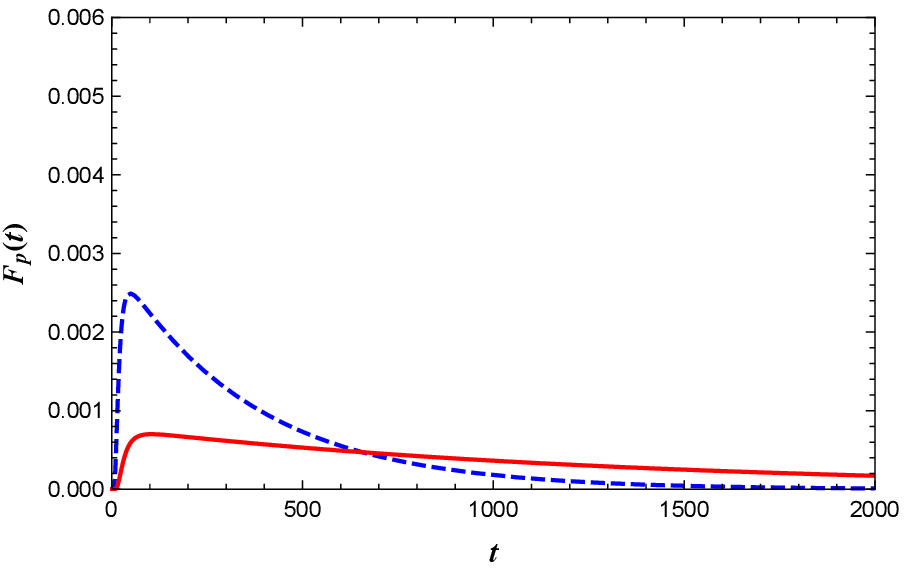}
\label{firstpasslbh}}
\caption{The probability distribution of the first passage time $F_p(t)$. Red solid and blue dashed curves are for the coexistent temperatures $T_E=0.0284$ and  $T_E=0.0307$, respectively, with $g=1$. (a) From SBH state to LBH state. (b) From LBH state to SBH state.}
\label{firstpass}
\end{figure*}

First we write the expression for the probability that the initial SBH state of black hole has not made a first passage by time $t$,
\begin{equation}
    \Sigma (t)=\int _0 ^{r_m}\rho (r,t) dt,
\end{equation}
where $r_m$ is obtained from the solution of the Eq. \ref{rmeq}. For the initial LBH state, the definition changes to,
\begin{equation}
    \Sigma (t)=\int _ {r_m} ^\infty \rho (r,t) dt,
\end{equation}
The probability distribution $\Sigma (t)$ is related to the first passage time $F_p(t)$ as,
\begin{equation}
    F_p(t)=-\frac{d\Sigma (t)}{dt}.
\end{equation}
For both initial states it can be easily shown that \citep{Li:2020nsy},
\begin{equation}
    F_p(t)=-D\frac{\partial }{\partial t}\rho (r,t)\Bigg|_{r=r_m},
\end{equation}
where the absorbing boundary condition at $r_m$ and reflecting boundary condition at the other end is imposed on the Fokker-Planck equation. The quantity $F_p(t)dt$ is the probability that the initial black hole state passes through the intermediate state located at $r_m$ for the first time in a time interval of $(t,t+dt)$. In the first passage, the black hole leaves the initial state and hence $\Sigma (r,t)| _{t\rightarrow \infty}=0$. The normalisation of the probability distribution is not preserved here.

We solve the Fokker-Planck equation numerically by imposing an absorbing boundary condition at $r=r_m$. As the Gibbs free energy is diverging at $r=0$, we take the reflecting boundary at $r=\epsilon$, where $\epsilon >0$.
The results are presented in Fig. \ref{H_time_evolution_absorbing}. We considered the initial state in both SBH and LBH for different temperatures. It is clear that the wave packets decays quickly as time passes. For a more intuitive understanding, we consider the profile of $\rho (r_{hi},t)$ from these plots for both the initial states, as shown in Fig. \ref{prob}. It is clear that the increase in temperature makes the probability drop faster. An important point to note here is that the probability is not conserved.

The corresponding distributions of the first passage time $F_p(t)$ are shown in Fig. \ref{firstpass}. The plots show a common behaviour for initial SBH and LBH states. In each case, there exists a peak near $t=0$, indicating a quick increase in the first passage time. This can be understood as a large number of first passage events occur in a short interval of time, and the distribution decays exponentially with time evolution. An increase in temperature makes it easy for the events to takes place, and as a result, the peak of the distribution of $F_p(t)$ increases, become sharp and shifted towards $t=0$.

We observe that the dynamics of phase transition of Hayward AdS black hole mimics that of charged AdS black hole, which is studied in Ref. \citep{Li:2020nsy}. In the next section, we will study the dynamics of one more regular black hole solution, namely Bardeen AdS black hole, that arises from the generic class of regular solutions that we are interested in. The observation will enable us to generalise the results for that class of regular black holes due to the coupling of non-linear electrodynamics.

\section{Bardeen AdS Black Hole}
\label{bardeensecion}
\subsection{Thermodynamics and phase transition}
In this section, we consider the Bardeen AdS spacetime which can be obtained from the following Lagrangian density, 
\begin{eqnarray}
\label{lagranb}
\mathcal{L\left(F\right)}=\frac{12}{\alpha}\frac{\left(\alpha \mathcal{F}\right)^{5/4}}{\left(1+\left(\alpha \mathcal{F}\right)^{1/2}\right)^{5/2}}.
\end{eqnarray}
The metric has the following form \citep{Fan:2016hvf},
\begin{equation}
ds^2=-f(r)dt^2+\frac{1}{f(r)}dr^2+r^2d\Omega ^2,
\end{equation}
with the metric function,
\begin{equation}
f(r)=1-\frac{2 M r^2}{\left(g^2+r^2\right)^{3/2}}+\frac{8}{3} \pi  P r^2.
\end{equation}
Where $g$ is the monopole charge parameter. The mass of the black hole is given by,
\begin{equation}
M=\frac{\left(g^2+r_+^2\right)^{3/2} \left(8 \pi  P r_+^2+3\right)}{6 r_+^2}.
\end{equation}
The Hawking temperature can be obtained from the surface gravity, which has the following form,
\begin{align}
T=\frac{2 P r_+^3}{g^2+r_+^2}+\frac{r_+}{4 \pi  \left(g^2+r_+^2\right)}-\frac{g^2}{2 \pi  r_+ \left(g^2+r_+^2\right)}.
\label{bardeentemp}
\end{align}
The first law of thermodynamics is identical to Hayward AdS spacetime,
\begin{equation}
dM=TdS+\Psi dQ_m+VdP+\Pi d \alpha,
\end{equation}
with the variables having the same meaning. Indeed, this is the generic form of the first law of black hole thermodynamics in the extended phase space for spacetimes with non-linear electric/magnetic charges, which can be obtained using a covariant approach \citep{Zhang:2016ilt}. The entropy of the black hole has the nontrivial form,
\begin{equation}
S=\int \frac{dM}{T}=-\frac{2 \pi  g^3 }{r_+}\, _2F_1\left(-\frac{3}{2},-\frac{1}{2};\frac{1}{2};-\frac{r_+^2}{g^2}\right),
\end{equation}
where $\, _2F_1$ is the Hyper-geometric function. The volume $V$ can be obtained from the first law as,
\begin{equation}
    V=\left( \frac{\partial M}{\partial P}\right)_{S,Q_m,\alpha}=\frac{4}{3} \pi  \left(g^2+r_+^2\right)^{3/2}.
\end{equation}
The equation of state for the system can be obtained from the Hawking temperature (Eq. \ref{bardeentemp}) as,
\begin{equation}
P=\frac{g^2}{4 \pi  r_+^4}+\frac{g^2 T}{2 r_+^3}-\frac{1}{8 \pi  r_+^2}+\frac{T}{2 r_+}.
\end{equation}
Like Hayward case, Bardeen AdS black hole shows a first order vdW phase transition between the LBH and SBH phase  \citep{Tzikas:2018cvs}. The critical values associated with this critical behaviour are given by,
\begin{equation}
T_{cB}=-\frac{\left(\sqrt{273}-17\right) \sqrt{\frac{1}{2} \left(\sqrt{273}+15\right)}}{24 \pi  g},
\end{equation}
\begin{equation}
P_{cB}=\frac{\sqrt{273}+27}{12 \left(\sqrt{273}+15\right)^2 \pi  g^2}.
\end{equation}
As in the previous case, the critical values depend on the parameter $g$.

\begin{figure*}[tbh]
\centering
\subfigure[ref2][]{\includegraphics[scale=0.8]{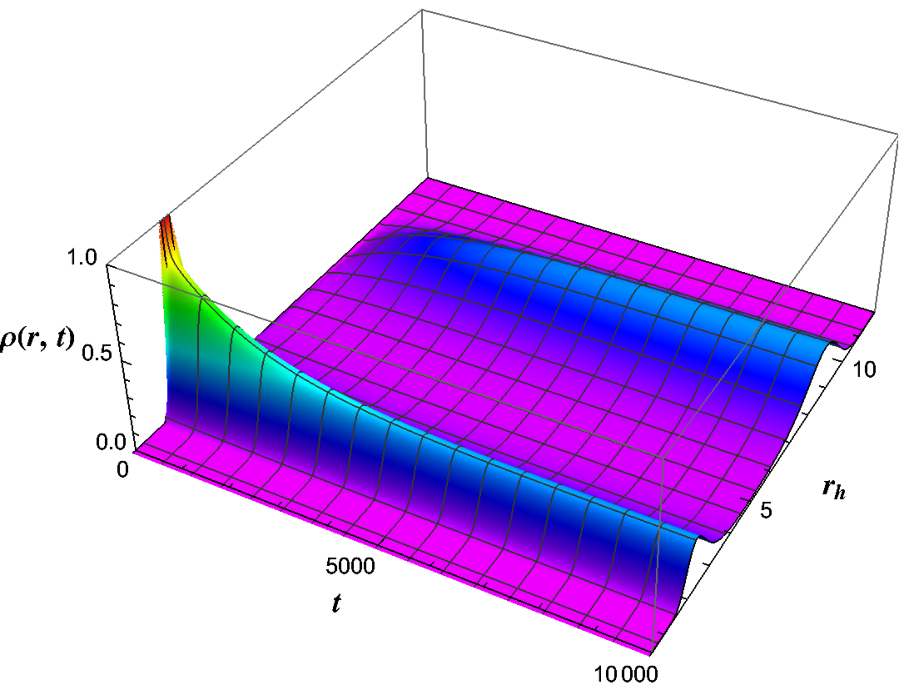}
\label{Brbcsbh3dt1}}
\qquad
\subfigure[ref2][]{\includegraphics[scale=0.8]{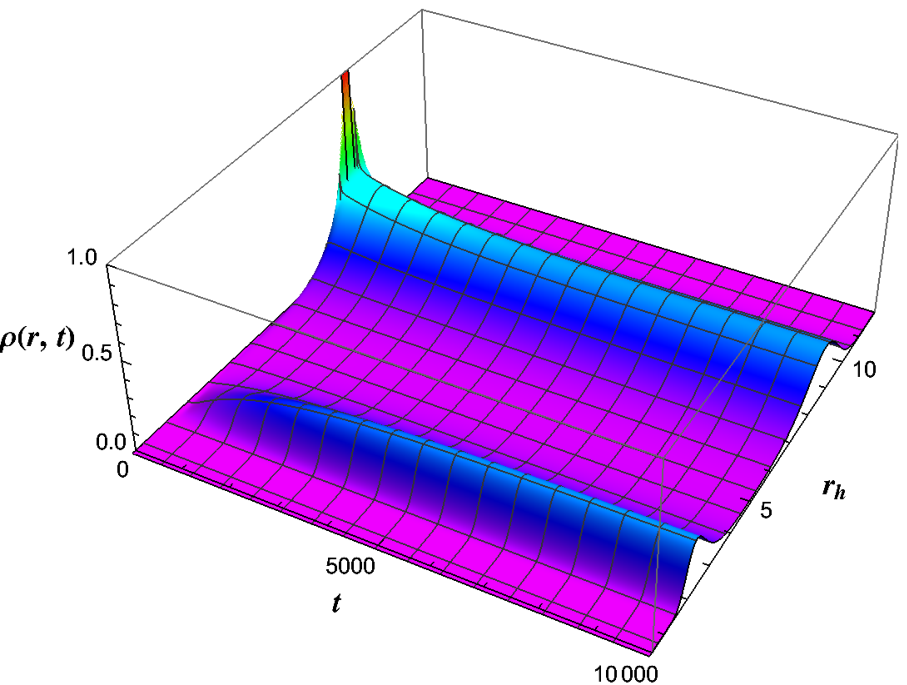}
\label{Brbclbh3dt1}}
\caption{The time evolution of the probability distribution $\rho (r, t)$ for Bardeen AdS black hole. (a) The initial Gaussian wave packet at SBH state, (b) initial wave packet at LBH state. The reflection boundary conditions are imposed at $r=0$ and $r=\infty$. The coexistent temperatures correspond to $P=0.6P_{cB}$ ($T_E=0.02077$ , with $g=1$.}
\label{B_time_evolution}
\end{figure*}

\begin{figure*}[tbh]
\centering
\subfigure[ref2][]{\includegraphics[scale=0.8]{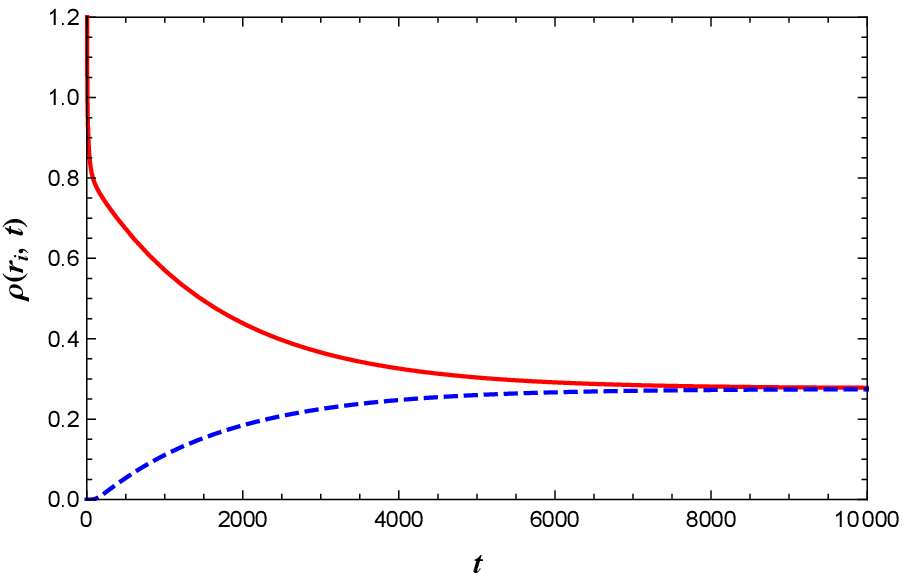}
\label{Brbcsbh2dt1}}
\qquad
\subfigure[ref3][]{\includegraphics[scale=0.8]{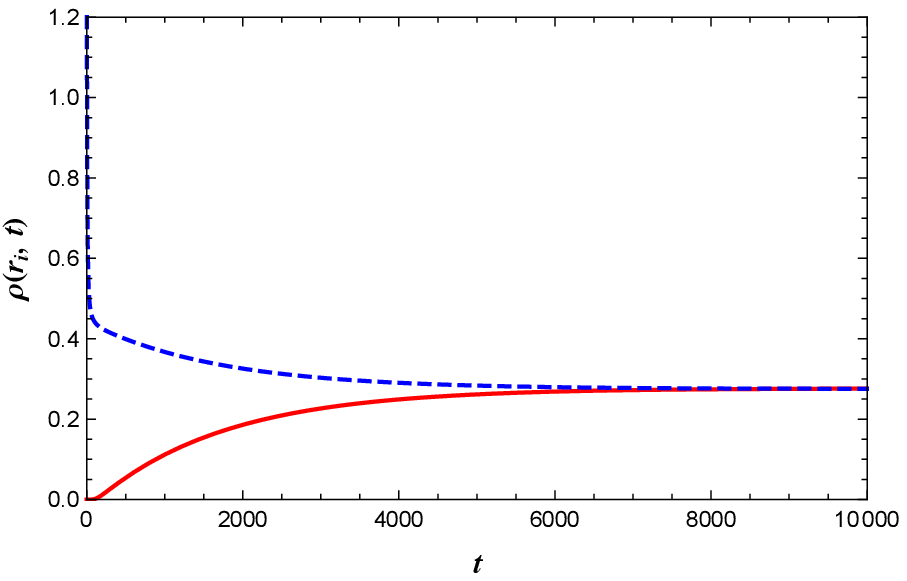}
\label{Brbcsbh2dt2}}
\caption{The time evolution of the probability distribution function $\rho (r,t)$ for Bardeen AdS black hole. The solid red and dashed blue curves correspond to $\rho (r_{hs},t)$ and $\rho (r_{hl},t)$, respectively. The initial wave packet is located at SBH state in (a) and at LBH state in (b). The coexistent temperatures is $T_E=0.02077$, with $g=1$.}
\label{B_time_evolution_2d}
\end{figure*}

\begin{figure*}[tbh]
\centering
\subfigure[ref2][]{\includegraphics[scale=0.8]{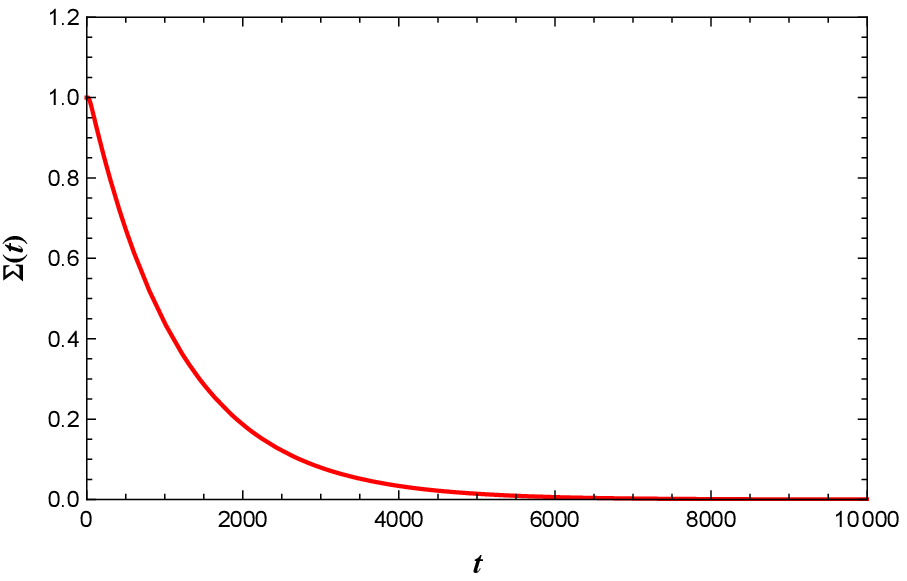}
\label{Bprobsbh}}
\qquad
\subfigure[ref3][]{\includegraphics[scale=0.8]{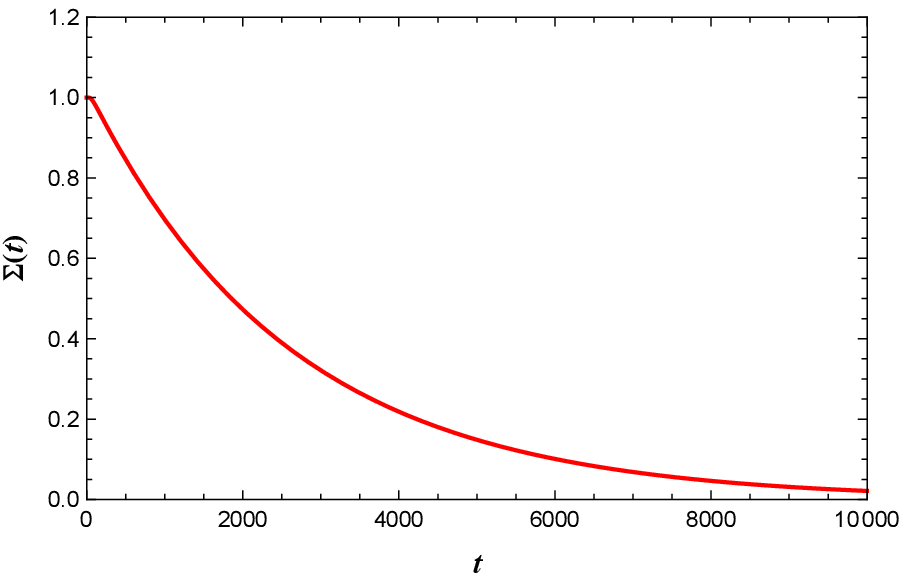}
\label{Bproblbh}}
\caption{The time evolution of the probability distribution $\Sigma (t)$ that the system stays at the initial state for Bardeen AdS black hole. (a) Initial SBH and (b) initial LBH state (b). We chose the coexistent temperature $T_E=0.02077$, with $g=1$.}
\label{Bprob}
\end{figure*}

\begin{figure*}[tbh]
\centering
\subfigure[ref2][]{\includegraphics[scale=0.8]{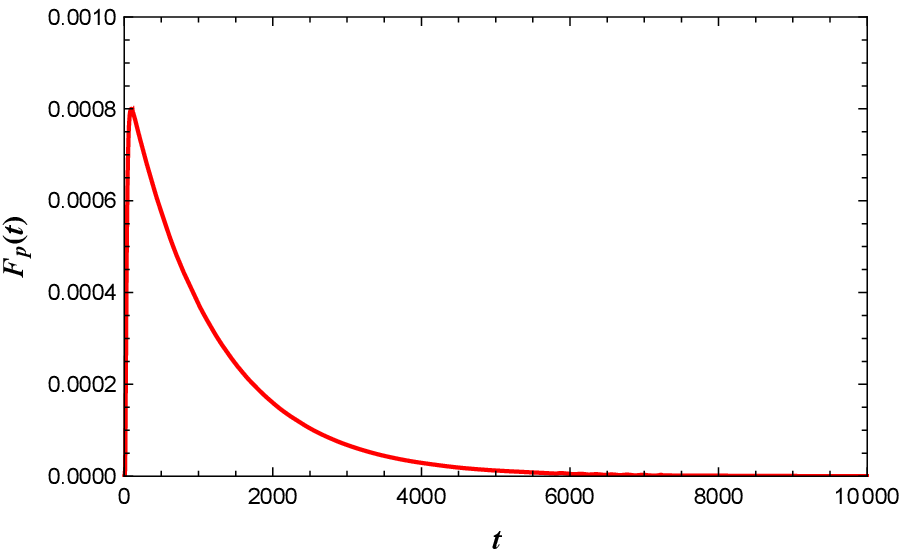}
\label{Bfirstpasssbh}}
\qquad
\subfigure[ref3][]{\includegraphics[scale=0.8]{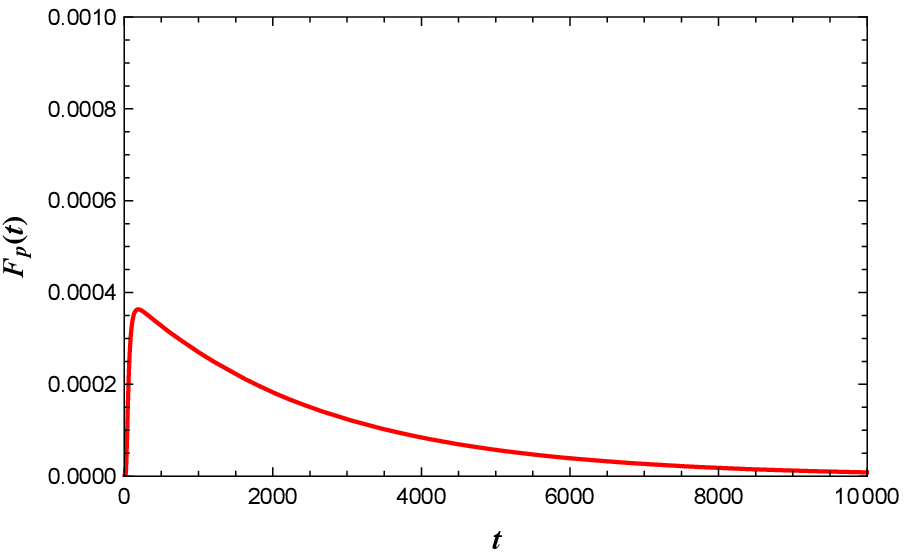}
\label{Bfirstpasslbh}}
\caption{The probability distribution of the first passage time $F_p(t)$ for Bardeen AdS black hole. We chose the coexistent temperature $T_E=0.02077$, with $g=1$. (a) From SBH state to LBH state. (b) From LBH state to SBH state.}
\label{Bfirstpass}
\end{figure*}

\subsection{Gibbs free energy landscape and dynamic properties of phase transition}
We carry out similar studies for Bardeen AdS black hole to investigate the dynamics of phase transition in the Gibbs free energy landscape. We find that the dynamical properties of the Bardeen case are qualitatively similar to the Hayward case. However, we note that there are quantitative differences between these two solutions. The time scale involved in the Bardeen case is much larger than the Hayward case. The results for a coexistent temperature corresponding to the pressure $P=0.6P_{cB}$ are shown in Figs. \ref{B_time_evolution}-\ref{Bfirstpass}. The effect of temperature on the dynamic process same as in the previous case.

\section{Discussions}
\label{discussions}

We investigate the dynamic properties of regular AdS black hole phase transitions in the free energy landscape. Our focus is on probing the probability evolution using the Fokker-Planck equation for black hole solutions due to Einstein gravity coupled to non-linear electromagnetic fields. The interesting feature of these solutions is that they do not possess the physical singularity at the centre, which is replaced with a de-Sitter core.  Of particular interest, we considered the special cases of such regular solutions, namely Hayward and Bardeen spacetimes. 

First, we present the extended thermodynamics and small-large black hole transition in brief. In the absence of analytical expression for the coexistence curve, we followed a naive method to study the dynamics of the phase transition between two coexistent phases using the swallowtail Gibbs free energy behaviour. The stable and unstable states of the black hole are identified as the minima and maxima, respectively, of the off-shell Gibbs free energy $G_L$, when plotted against the order parameter horizon radius $r_h$. The coexistent condition is identified with the equal depth wells of the free energy. 

The Fokker-Planck equation is solved numerically by imposing reflecting boundary conditions at $r=0$ and $r=\infty$ and with an initial Gaussian wave packet. The solutions are sought for initial wave packets at small and large black hole state at different temperatures. The results show that the initial black hole state will make a transition to other state as time evolves. The probability distribution saturates on both states after a long time, and the temperature guides this saturation time; the higher the ensemble temperature quicker it relaxes. In the next part of our investigation,  the first passage time for the process is studied in detail. Here, the Fokker-Planck equation is solved numerically by imposing the absorbing boundary condition at the unstable state, which is the location of the peak of the potential barrier. The peak in the distribution hints at a large number of first passage events occurring in a short period of time. The effect of temperature is studied in this case too, which shows that the occurrence of the events is favoured by higher temperature. 

The key idea in analysing the black hole system using free energy landscape is that, being a thermal system, it must be an emergent macroscopic state of the underlying spacetime degrees of freedom. With this inspiration, we believe that our study provides more insights into the microscopic structure of the black hole in the absence of central singularity. This study also reveals the proximity of regular solutions to charged AdS black holes. We expect that other regular black hole solutions will also exhibit similar microscopic degrees of freedom.



  \bibliography{BibTex}

\end{document}